# Comparison of the Baseline and the Minimal Steel Yoke Superconducting Magnets for the Future Circular Hadron-Hadron Collider

V. I. Klyukhin[1,2], A. Ball[2], C. P. Berriaud[3], E. Bielert[2], B. Curé[2], A. Dudarev[2], A. Gaddi[2], H. Gerwig[2], A. Hervé[4], M. Mentink[2], H. Pais Da Silva[2], U. Wagner[2], H. H. J. Ten Kate[2]

[1]*Skobeltsyn Institute of Nuclear Physics, Lomonosov Moscow State University, RU-119991, Moscow, Russia*

[2]*CERN, CH-1211, Geneva 23, Switzerland*

[3]*CEA Irfu, Saclay, 91191 France*

[4]*University of Wisconsin, Madison, WI 53706, USA*

Contact e-mail: *vyacheslav.klyukhin@cern.ch*

**Abstract** The conceptual design study of a hadron Future Circular hadron-hadron Collider (FCC-hh) with a center-of-mass energy of the order of 100 TeV assumes using in the experimental detector the superconducting magnetic system with a central magnetic flux density of an order of 4 T. A superconducting magnet with a minimal steel yoke was proposed as an alternative to the baseline iron-free solenoids. In a present study, both designs are modeled with Cobham's program TOSCA and compared. All the main parameters are discussed.

***Keywords*** Superconducting solenoid

## 1 Introduction

The conceptual design report study of the Future Circular hadron-hadron Collider (FCC-hh) [1] with a center-of-mass energy of the order of 100 TeV, assumed to be constructed in a new tunnel of 80–100 km circumference, is to be presented at the end of 2018. As a part of this study, a conception of a detector for hadron-hadron physics is developing. The FCC-hh detector comprises an iron-free magnetic system that consists of three superconducting solenoids: the main coil with a central magnetic flux density of 4 T and two auxiliary forward coils with a central magnetic flux density of 3.2 T each [2].

Following the recent alternative study of the magnetic system for the FCC-hh detector based on the superconducting solenoid with a minimal steel yoke [3], a possibility of using the steel yoke in the same baseline layout of the coils and the particle sub-detectors [4] is considered in present paper, and the main parameters of both magnetic systems are compared. For these comparisons, both, the baseline FCC-hh detector magnetic system [2], and the magnetic system with the minimal steel yoke, are calculated with a program TOSCA from Cobham CTS Limited [5].

## 2 Modelling the Magnetic Systems

The FCC-hh detector baseline magnetic system design, shown in Fig. 1, includes three components: the main superconducting coil with a total current of 69.6 MA-turns, and two superconducting forward coils with a total current of 12.6 MA-turns each.

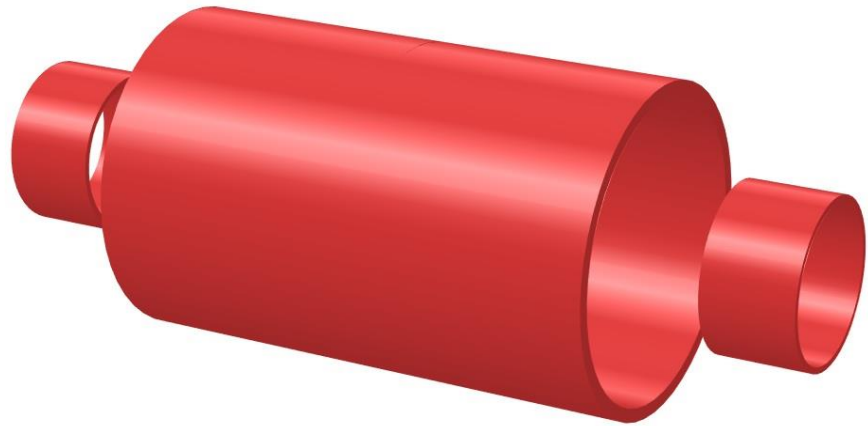

**Fig. 1** 3-D model of the FCC-hh detector baseline magnetic system comprising the main superconducting coil with 10.9 m inner diameter, and two superconducting forward coils with 5.6 m inner diameter. The distances between the coils are of 2.823 m each side.

In the calculations, the radial thickness of the main coil is assumed of 0.5 m, and the radial thickness of each forward coil is of 0.23 m. The distances between the main and each forward coils are of 2.823 m. Preparing the design of the baseline magnetic system this distance is chosen as a compromise between the value of the magnetic flux density in the transition region between the coils, and the attractive axial force onto the forward coil that is minimized to 61.8 MN. The compression axial force in the main coil middle plane is of 610.7 MN that creates a pressure of 34.1 MPa. The radial pressure in the main coil is of 1.55 MPa, and in each forward coil of 2.66 MPa. The stored energy of the baseline magnetic system is of 13.8 GJ.

To realize the minimal steel yoke conception [3], the coil is surrounded with five barrel wheels of 17.7 m outer diameter and 3.9 m width each as is shown in

Fig. 2. Each barrel wheel has two steel layers of 0.75 m thick with a radial gap of 0.5 m between them.

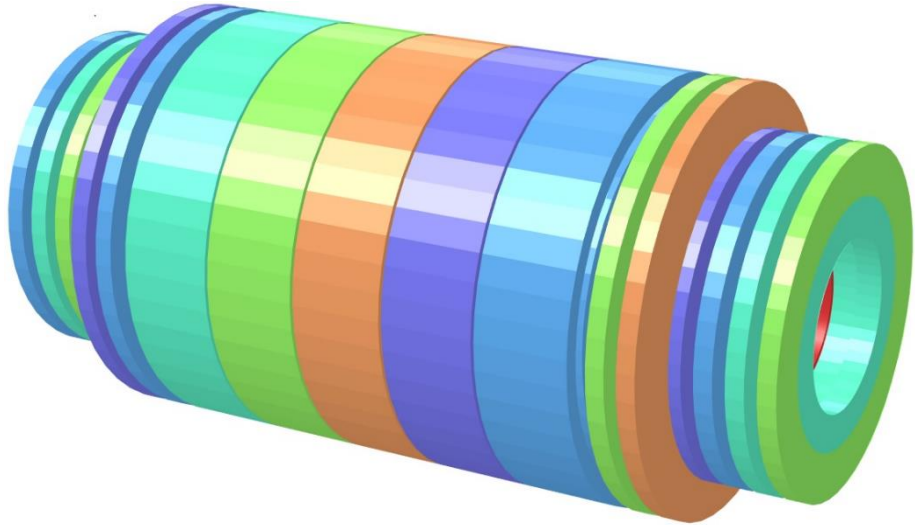

**Fig. 2** 3-D model of the FCC-hh detector minimal yoke magnetic system viewed from outside.

At the distances of 0.75 m off the extreme wheels, two end-cap disks of 17.7 m outer diameter and 0.75 m thickness are located at each barrel end. Four smaller disks of 14 m outer diameter at each yoke end follow these disks. The thicknesses of the first small disks is 0.5 m, other disks have the thicknesses of 0.75 m.

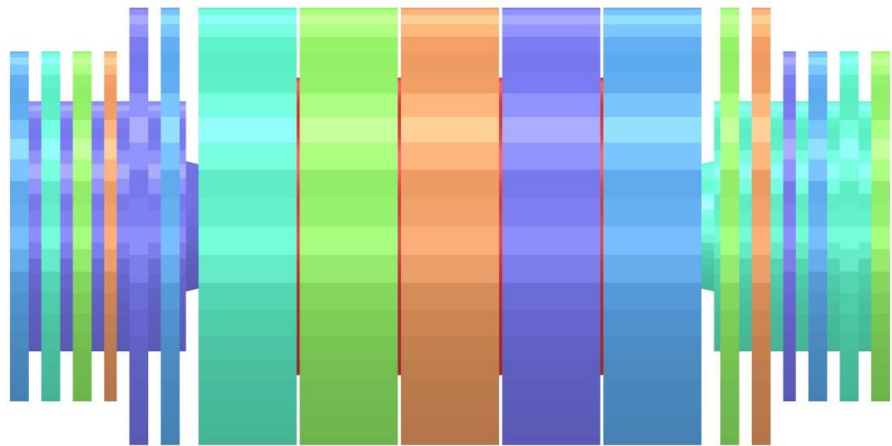

**Fig. 3** The main superconducting coil of 11.9 m outer diameter, the five barrel wheels of 3.9 m width each, the two conical-cylindrical shields of 10 m outer diameter each, the four end-cap disks of 17.7 m diameter each, and the eight end-cap disks of 14 m diameter each. The solenoid coil is visible between the barrel wheels in the air gaps of 0.125 m each. The length of the barrel part is 20 m; the total length of the yoke is 35 m

The air gaps of 0.5 m between all the end-cap disks serve for an installation of the muon detection chambers located in the same positions as in the baseline design of the FCC-hh detector [4]. As shown in Fig. 3, the inner parts of the end-cap disks rely on the cylindrical radiation protection shielding of 1 m radial thickness assumed to be made of a carbon steel. Inside the main coil this shield has a conical shape following the polar angles corresponded to the pseudorapidity region from ±2.1 to ±2.5 [4]. In opposite, in the baseline design of the magnetic system this shield is assumed to be made of non-magnetic material. The total length of the minimal steel yoke is 35 m.

## 3 Comparison of the Magnetic Systems

In Figs. 4 and 5, the magnetic flux density distribution is displayed in a vertical YZ-plane of the baseline and minimal steel yoke magnetic systems, respectively.

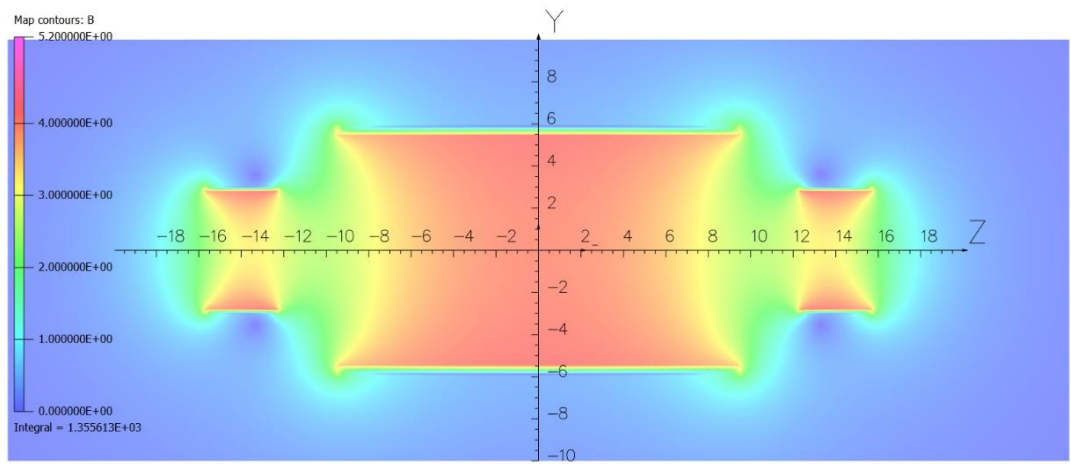

**Fig. 4** Magnetic flux density distribution in a vertical plane of the FCC-hh detector baseline magnetic system. The color magnetic field map plotted with the cell size of 0.05 m has the width of 50 m and the height of 20 m. The color scale unit is 1 T. The minimum and maximum magnetic flux density values are 0.0142 and 4.1595 T.

The magnetic flux density in the center of the main coil is 4 T in the baseline and 4.26 T in the minimal steel yoke magnetic systems, accordingly. As displayed in Fig. 6, the contribution of the steel yoke in the central magnetic flux density at the level of 6.5 % is compensating by the decreasing the magnetic flux density at the axis of the forward coils. At the distances of ±13.53 m from the main coil center, it is 3.2 T in the baseline and 3.07 T (4 % lower) in the minimal steel yoke magnetic systems, respectively. Fig. 5 shows that the magnetic flux density in the conical parts of the shielding is extremely high and in any design this part should be made of a non-magnetic material. Being ferromagnetic, this part creates enormous attractive axial force on the shield of 131 MN. This gives the main contribution into compression axial force onto the central barrel wheel of 240.5 MN and into attractive axial force onto each end-cap of 168.7 MN.

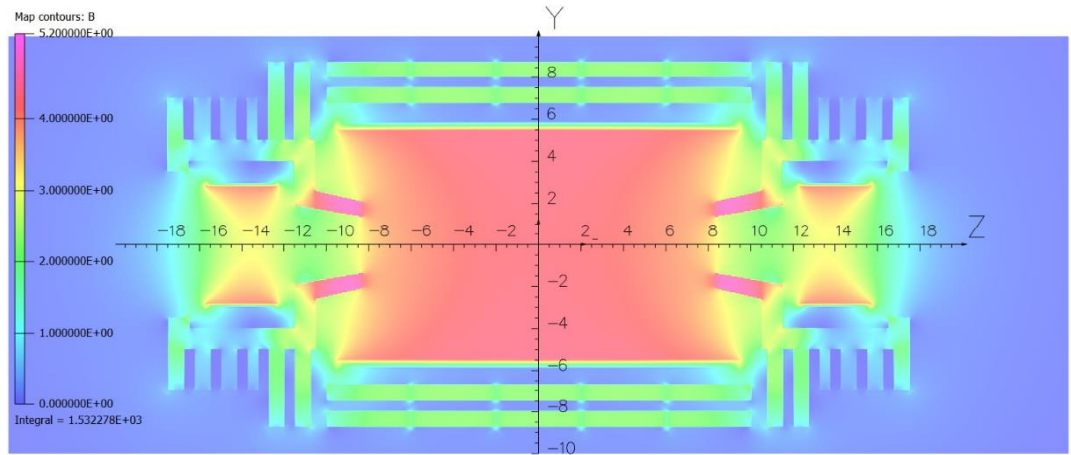

**Fig. 5** Magnetic flux density distribution in a vertical plane of the FCC-hh detector minimal yoke magnetic system. The color magnetic field map plotted with the cell size of 0.05 m has the width of 50 m and the height of 20 m. The color scale unit is 1 T. The minimum and maximum magnetic flux density values are 0.0002 and 5.1425 T. The maximum magnetic flux density in the barrel wheel layers is 2.3 T. The magnetic flux density in the first end-cap disks at the radius of 6 m is 2.4 T.

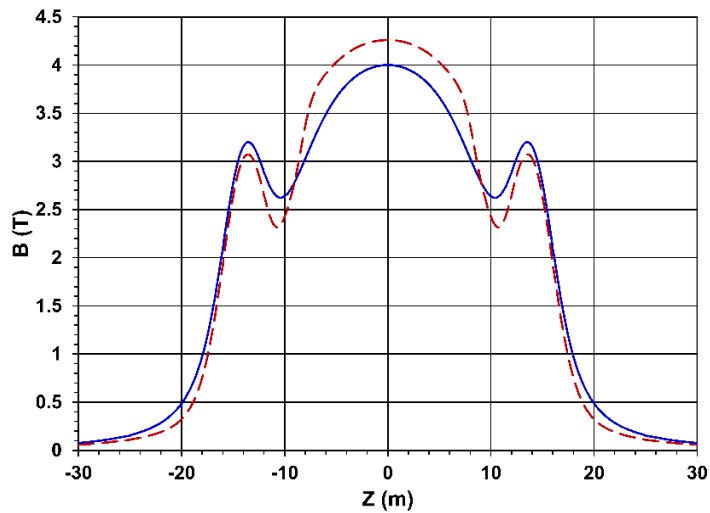

**Fig. 6** Magnetic flux density variation along the coil axis in the baseline (*smooth curve*) and minimal yoke (*dashed line*) magnetic systems.

The axial force on each forward coil in the minimal steel yoke magnetic system is of 66.6 MN. The compression axial force in the main coil middle plane is of 515.3 MN that creates a pressure of 28.8 MPa. The radial pressure in the main coil is of 2.49 MPa, and in each forward coil is of 2.28 MPa. The stored energy is of 14.6 GJ that is larger than in the baseline design.

Fig. 7 presents the stray magnetic flux density variation vs. a radius in the middle plane of the main coil, and vs. a distance from the main coil center along the coil axis. The stray magnetic flux density drops to the safety level of 5 mT at a radius of 59.4 m (46 m) from the coil axis in the baseline (minimal steel yoke) design. The stray field drops to 5 mT at a distance of 64.9 m (60.4 m) from the main coil center along the coil axis in the baseline (minimal steel yoke) design.

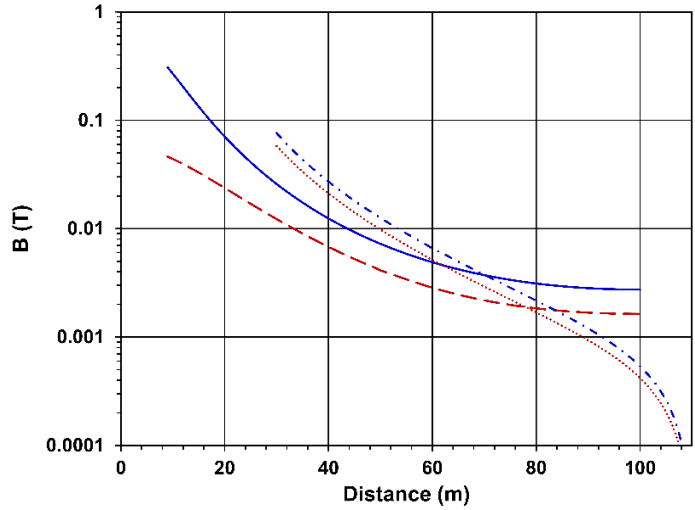

**Fig. 7** Magnetic flux density out of the coil in the main coil central plane vs. a radius (*smooth* and *dotted lines*) as well as along the coil axis vs. a distance from the main coil center (*dash-dotted* and *small dotted lines*). Smooth and dash-dotted lines correspond to the baseline magnetic system. Dashed and small dotted lines correspond to the minimal steel yoke magnetic system.

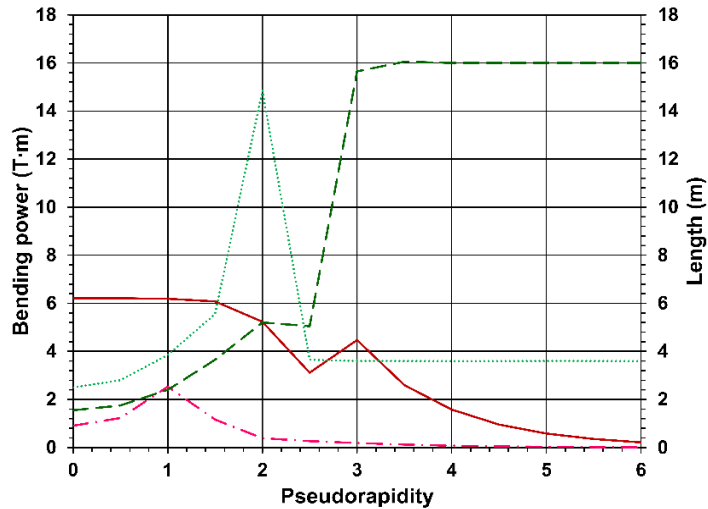

**Fig. 8** Magnetic flux density bending component integrals (*left scale*, *solid* and *dash-dotted lines*), and the length of the charged particle trajectory (*right scale*, *dashed* and *small dotted lines*) in the inner tracker (*smooth* and *dashed lines*), and in the muon system (*dash-dotted* and *small dotted lines*) of the baseline FCC-hh detector design vs. the pseudorapidity.

The last but not the least parameters for the comparisons are the integrals of the magnetic flux density bending component orthogonal to the charged particle trajectory vs. the pseudorapidity inside the inner tracker of 1.5563 m radius and 16 m length, and through the muon system. For both magnetic systems, the bending powers and the track lengths passing by the charged particles in the inner tracker and in the muon system are shown in Figs. 8 and 9.

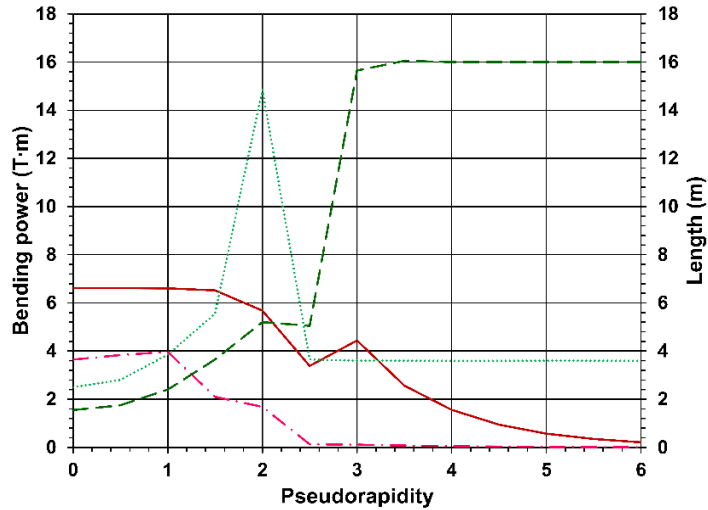

**Fig. 9** Magnetic flux density bending component integrals (*left scale*, *solid* and *dash-dotted lines*), and the length of the charged particle trajectory (*right scale*, *dashed* and *small dotted lines*) in the inner tracker (*smooth* and *dashed lines*), and in the muon system (*dash-dotted* and *small dotted lines*) of the FCC-hh detector with the minimal steel yoke vs. the pseudorapidity.

These plots finally display that the minimal steel yoke design has no substantial advantage in comparison with the baseline magnetic system.

## 4 Conclusions

This study investigates the advantages of the superconducting solenoid magnetic system with a minimal steel yoke in comparison with the baseline iron-free magnetic system for the detector at the Future hadron-hadron Circular Collider with a center-of-mass energy of 100 TeV. The compared parameters of two magnetic system confirm the choice in a favor of the baseline design with respect to the minimal steel yoke design. The latest is more expensive and has no substantial advantages in comparison with the FCC-hh detector baseline magnetic system.